\documentclass[10pt, conference, letterpaper]{IEEEtran}
\usepackage[utf8]{inputenc}
\usepackage[english]{babel}
\usepackage{ upgreek }
\usepackage[pdftex]{graphicx}
\usepackage{cite}
\usepackage{amsmath}
\usepackage{amssymb}
\usepackage{caption}
\usepackage{subcaption}
\usepackage{stfloats}
\usepackage{tikz}
\usepackage{todonotes}

\usepackage{tabularx}
\usepackage{multicol}
\usepackage{epstopdf}
\usepackage{enumitem}
\usepackage{url}
\usepackage{amsmath}

\graphicspath{{Pics/}}
\DeclareGraphicsExtensions{.eps,.pdf,.png,.jpg}

\IEEEoverridecommandlockouts

\usepackage{tikz}
\usepackage{textcomp}
\usepackage{hyperref}

\begin{document}
\bstctlcite{IEEEexample:BSTcontrol}

\title{Enabling Synchronous Uplink NOMA in Wi-Fi Networks
\thanks{The research was done at IITP RAS and supported by the Russian Science Foundation No 21-19-00846, https://rscf.ru/project/21-19-00846/}}

\author{\IEEEauthorblockN{
Grigory Korolev\IEEEauthorrefmark{1}\IEEEauthorrefmark{2},
Aleksey Kureev\IEEEauthorrefmark{1}\IEEEauthorrefmark{2}\IEEEauthorrefmark{3},
Evgeny Khorov\IEEEauthorrefmark{1}\IEEEauthorrefmark{2}\IEEEauthorrefmark{3}, and
Andrey Lyakhov\IEEEauthorrefmark{1}\IEEEauthorrefmark{2}
}
\IEEEauthorblockA{ 
\IEEEauthorrefmark{1}Moscow Institute of Physics and Technology, Moscow, Russia\\
	\IEEEauthorrefmark{2}Institute for Information Transmission Problems, Russian Academy of Sciences, Moscow, Russia\\
\IEEEauthorrefmark{3}National Research University Higher School of Economics, Moscow, Russia\\	
Email: \{korolev, kureev, khorov, lyakhov\}@wireless.iitp.ru}

\vspace{-1.5em}}

\maketitle 

\vspace{-2em}
\begin{abstract}
Non-Orthogonal Multiple Access (NOMA) is a promising technology for future Wi-Fi. In uplink NOMA, stations with different channel conditions transmit simultaneously at the same frequency by splitting the signal by power level. Since Wi-Fi uses random access, the implementation of uplink NOMA in Wi-Fi faces many challenges. The paper presents a data transmission mechanism in Wi-Fi networks that enables synchronous uplink NOMA, where multiple stations start data transmission to the access point simultaneously. The developed mechanism can work with the legacy Enhanced Distributed Channel Access (EDCA) mechanism in Wi-Fi. With simulation, it is shown that the developed mechanism can double the total throughput and geometric mean throughput compared with the legacy EDCA.

\end{abstract}

\section{Introduction} 
\label{sec:introduction}
Wi-Fi is the most popular, relatively cheap, and easily deployable wireless technology for data transmission \cite{cisco}. Wi-Fi is used in various scenarios with many user devices: airports, universities, malls, stadiums, etc. 
In this regard, the problem of the throughput increase becomes especially acute.

Non-orthogonal multiple access (NOMA) is a new approach to achieve throughput gain in Wi-Fi networks. 
NOMA allows devices to transmit multiple data flows at the same time on the same frequency.  In the downlink NOMA (DL-NOMA), the access point (AP) transmits data for several stations (STA), while in the uplink NOMA (UL-NOMA), multiple stations transmit frames to one AP.

While DL-NOMA demonstrates its effectiveness in Wi-Fi \cite{kureev2020prototyping, vaezi2019interplay}, the usefulness of the UL-NOMA in Wi-Fi is an open question.
To explain the challenges of UL-NOMA in Wi-Fi, let us describe how it works. 

Consider a Wi-Fi network with two STAs and one AP. 
The STAs are located at different distances from the AP and have various channel conditions. 
Both STAs transmit frames to the AP. 
In UL-NOMA, the AP receives a superposition of signals (referred to as a NOMA frame) from both STAs. 
We study a synchronous UL-NOMA, which assumes that the signals arrive at the AP simultaneously.
Successful reception of both frames is possible if the powers of the received signals from the STAs are different: the power from a far STA is lower than the power of a STA located closer to the AP. To decode such a signal, the AP uses the Successive Interference Cancellation (SIC) method. SIC allows the AP to decode signals sequentially. First, the AP decodes the stronger signal considering the other one as noise. 
Second, the AP subtracts the decoded signal from the original one and decodes the rest of the signal.

Although the idea of  UL-NOMA is quite simple, in practice, this method faces many implementation issues.

The main one is related to the organization of UL-NOMA transmission in Wi-Fi. The easiest way is to schedule simultaneous transmissions by various STAs.  However, scheduled transmission breaks the backward compatibility and makes UL-NOMA incompatible with Enhanced Distributed Channel Access (EDCA) traditionally used in Wi-Fi.

In this work, we develop and study a mechanism that enables synchronous UL-NOMA transmissions upon the legacy EDCA mechanism. The mechanism is based on well-known Request-to-Send/Clear-to-Send (RTS/CTS) method and uses reservation signals (RSs) for UL-NOMA transmission management. We evaluate the performance of the developed mechanism using simulation. The results show that the use of UL-NOMA in Wi-Fi provides up to 100\% gain in throughput compared to EDCA.

The rest of the work is organized as follows. Section~\ref{sec:related} gives an overview of papers exploring the NOMA. Section~\ref{sec:protocol} describes the data transmission mechanism,  and Section~\ref{sec:scenario} explains the scenario of the experiment and the formulation of the problem. The results are discussed in Section~\ref{sec:results}. Section~\ref{sec:conclusion} concludes the paper.

\section{Related Works} 
\label{sec:related}
UL-NOMA is widely used in 5G cellular systems~\cite{al2014uplink, gerasin2020dynamic}. Since they mainly use scheduled channel access, synchronous UL-NOMA perfectly fits the cellular network architecture. 
Thus, the base station knows about the channel conditions of all users and can organize a UL-NOMA transmission~\cite{tabassum2016non}.
However, there is no way in Wi-Fi to do the same if STAs use random channel access.

In our previous paper~\cite{korolev2020performance}, we have considered an asynchronous NOMA transmission mechanism in Wi-Fi that can be used with EDCA.  
In the proposed scheme, the primary frame is divided into NOMA slots. 
In each NOMA slot, Secondary STA competes for the channel access.
The Secondary STA starts transmission in a successful NOMA slot within one primary frame. 
However, this mechanism has implementation issues connected with independent channel estimation for all frames in the NOMA frame. 

Several works consider a synchronous NOMA transmission in Wi-Fi.
Paper~\cite{ghazi2018uplink} consider the transmission mechanism that is based on Channel State Sorting-Pairing Algorithm (CSS-PA). 
The AP chooses the two STAs based on their channel condition and schedules a UL-NOMA transmission for this pair.
The paper \cite{pavan2021novel} provides an UL-NOMA transmission mechanism that works with EDCA.
The AP groups the STAs for UL-NOMA transmission,  depending on their Modulation and Coding Schemes (MCS).
Each group has a head STA that is closer to the AP than the others.
In the proposed scheme, only head STAs contend for the channel, while the other STAs transmit data only in UL-NOMA. 
This feature limits the rest STAs transmission, which leads to unfairness in channel resource consumption.

In the next section, we develop a synchronous UL-NOMA transmission mechanism that works with EDCA and allows all STAs to transmit in EDCA with UL-NOMA.

\section{Transmission mechanism} 
\label{sec:protocol}
\subsection{EDCA with RTS/CTS}
Since the developed mechanism is based on EDCA, let us describe the basic EDCA rules.

In EDCA, every STA independently decides when to start a transmission. Specifically, after every transmission attempt, the STA initializes a backoff counter and waits for a random number of slots with duration $\sigma$ each. This random number is in the range $[0, CW-1]$, where CW is a contention window.
The STA decrements the backoff counter each time when the medium is idle during the slot time $\sigma$. If the medium becomes busy, the STA suspends the backoff counter until the medium is idle again during the Arbitration InterFrame Space (AIFS) interval. The STA starts the transmission when the backoff counter reaches zero.
If several STAs start transmissions at the same time, the collision occurs, and frames are corrupted.
When the AP successfully decodes the transmitted data, it responds with an Acknowledgement frame (ACK) after the Short InterFrame Space (SIFS) interval.
If the RTS/CTS mechanism is enabled, the STA starts the transmission with an RTS frame. 
In this case, the AP responds with a CTS frame.
Any other STAs that receive RTS or CTS consider the channel as busy for the time specified in these frames. 




\subsection{NOMA RS}
\begin{figure}[h!]
	\centering{
		\includegraphics[width = 0.9\linewidth]{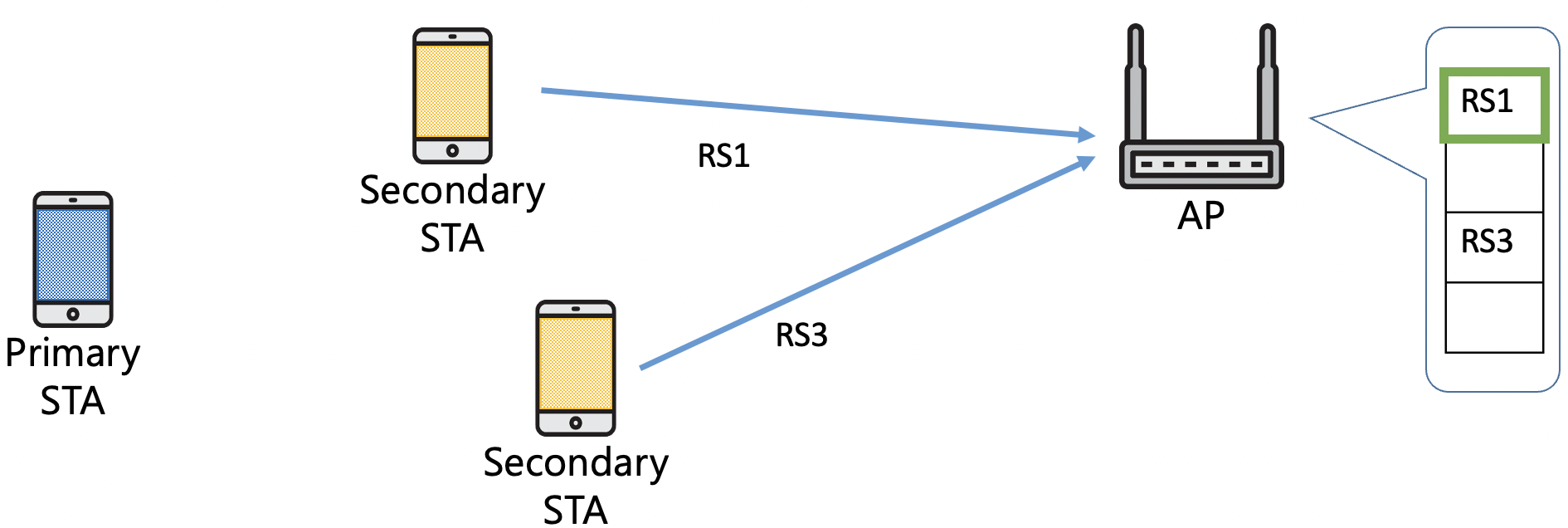}
		\caption{An example of UL-NOMA scenario.}
		\label{figure:general_view}
	}
\end{figure}

\begin{figure}[h!]
\centering{
	\includegraphics[width = 0.98\linewidth]{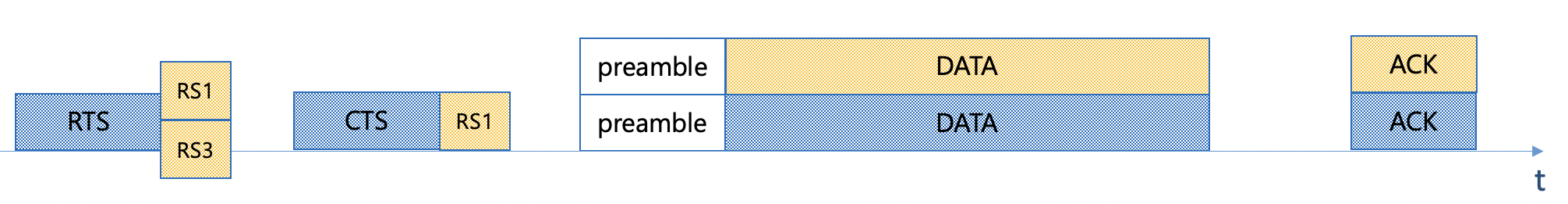}
	\caption{The structure of transmitting frames.}
	\label{figure:frames}
}
\end{figure}
We call the developed mechanism NOMA RS.
It starts from the successful EDCA transmission using RTS/CTS.

Denote the STA that performs a successful EDCA transmission as a Primary STA, and the STA that transmits with UL-NOMA over the primary frame as a Secondary STA. 

The transmission mechanism has the following features:

\begin{itemize}
\item  It uses RTS/CTS and RS mechanisms together with EDCA.
\item  It works upon EDCA without breaking the random access principles.
\item  It reduces the number of collisions of frames from Secondary STAs.
\end{itemize}


In addition to the frame duration, the preamble of the successful primary frame from Primary STA contains the information about the MCS used, which is dynamically selected based on the Signal-to-Noise Ratio (SNR). Therefore, with the help of MCS, it is possible to estimate how far the transmitting STA is located from the AP.

We assume perfect knowledge of the frame power at the receiving device from the Primary STA. 
If the received frame power is $ P_ {RX}^{i} $ and the noise power is $ P_ {n} $, then
\begin{equation}
SNR_{i} = \frac{P_{RX}^i}{P_n}.
\label{eq:leg_SNR}
\end{equation}

This information is evaluated during the previous data transmission between $ STA_{i} $ and the AP.
After reading the LTF (Long Training Field) that contains information about $ SNR_ {i} $, $ STA_ {j} $ evaluates its own $ SNR_ {j}^{NOMA} $, considering that there will be a transmission with NOMA:

\begin{equation}
SNR_j^{NOMA} = \frac{P_{RX}^j}{P_n + P_{RX}^i}.
\label{eq:NOMA_SNR}
\end{equation}

$ STA_j $ decides to use NOMA if all the following conditions are met:

\begin{enumerate}
\item $ STA_j $ is within the coverage of $ STA_i $;
\item $ SNR_j ^ {NOMA} \geq \gamma_ {min} $,
\label {eq: NOMA_condition}
\end {enumerate}
\noindent where $ \gamma_ {min} $ is the minimum SNR threshold that allows a STA to use the most robust MCS with index 0.

The proposed mechanism only works if the Primary STA uses RTS/CTS. When the Primary STA transmits an RTS frame to the AP, $ STA_j $ realizes that it may transmit as the Secondary STA and sends its unique orthogonal reservation signal $ RS_j $ to the AP. 
The RS can be any signal with a duration of 4 $\mu$s, which can be orthogonalized between multiple STAs.
For instance, the RS can be a set of subcarriers, which is used to transmit the OFDM signal.
In this work, we assume that the frequency band occupied by the RS is equal to the bandwidth of the minimum resource unit in IEEE 802.11ax networks i.e., 78.125 $kHz$.  If the number of RSs exceeds the number of STAs in the Wi-Fi network, all RSs are considered to be unique for each STA in this network.

The AP allocates the RS to the STA during the association handshake.  Thus, according to the received RSs, the AP selects the Secondary STA, see Figs. \ref{figure:general_view} and \ref{figure:frames}, where 2 STAs send their RSs and the AP selects the STA with RS1 as a Secondary one. Along with the CTS for the Primary STA, the AP also sends the selected RS, which is only processed by the specified Secondary STA. Therefore, a NOMA transmission is initiated. 

The AP chooses the strategy to select a particular STA as the Secondary STA, e.g., the AP can maximize the overall throughput in the network or allocate channel resources in a proportional fair manner.  
Let $ r_i $ be the data rate of the STA $i$ defined by the current MCS without the usage of NOMA  and $ R_i $ be the throughput of the STA $i$ measured during some large time interval before the current transmission. 

We consider the following strategies for selecting Secondary STAs:

\begin {itemize}
\item MaxRate that maximizes throughput and uses the $ r_i $ as the metric;
\item Proportional Fair that aims to maximize throughput provided that the resources are allocated fairly and uses $ \frac {r_i} {R_i} $ as the metric.
\end {itemize}
With both strategies, the AP selects a STA, the metric of which is higher.

In case of a successful NOMA transmission, the AP decodes both signals (from the Primary and Secondary STA) and sends an acknowledgment frame using DL-NOMA.
Since the Primary STA has worse channel conditions than the Secondary STA, the downlink NOMA-ACK frame consists of a low-power ACK for the Secondary STA and a high-power ACK for the Primary STA, i.e., the AP uses the DL-NOMA.

\section{Scenario} 
\label{sec:scenario}
\begin{table}[h!]
	\caption{SNR thresholds for various MCS.}
	\centering
	\begin{tabular}{|| c | c | c ||} 
		\hline
		MCS index & Bitrate (Mbps) & SNR (dB) \\ [0.5ex] 
		\hline\hline
		0 & 8.6 & 3.9 \\
		1 & 17.2 & 6.9 \\
		2 & 25.8 & 9.9 \\
		3 & 34.4 & 13.5 \\
		4 & 51.6 & 16.6 \\
		5 & 68.8 & 21.4 \\
		6 & 77.4 & 22.6 \\
		7 & 86.0 & 23.8 \\
		8 & 103.2 & 28.5 \\
		9 & 114.7 & 29.7 \\
		10 & 129.0 & 33.2 \\
		11 & 143.4 & 35.1 \\
		\hline
	\end{tabular}
	\label{tab:MCS}
\end{table}

All STAs operate in the saturation mode, i.e., the transmission queue of each STA is assumed to be always nonempty. 

Consider a Wi-Fi network with an AP located in the center of a circle of radius $R$ and $ N $ STAs uniformly distributed within the circle. We consider a small radius $R=R_1$ such that each STA can reliably receive data at MCS0 from any STA, and a large radius $R=R_2$ that gives us hidden STAs. 

By default, all STAs use EDCA and transmit data frames with a fixed payload $l$. In this paper, we use the Log-distance propagation model \cite{srinivasa2009path}, which determines the loss $PL$ as follows:
\begin{equation}
	PL = P_{TX} - P_{RX} = PL_0 + 10 \beta \log_{10} \frac{d}{d_0}, 
\end{equation}   
\noindent where $ PL_0 $ is the loss at the distance $ d_0 $, $ \beta $ is the loss exponent, $ P_ {TX} $ is the transmission power of each STA together with the antenna gain.
We assume that the transmission powers of all STAs are the same.
Before transmission, each STA selects the MCS based on the SNR obtained with (\ref{eq:leg_SNR}) for a Primary STA and with (\ref{eq:NOMA_SNR}) for a Secondary STA.
The MCS is selected from Table \ref{tab:MCS}, which contains the reference thresholds for the IEEE 802.11ax standard \cite{khorov2018tutorial}. 

\begin{table}[h!]
	\caption{Simulation parameters.}
	\centering
	\begin{tabular}{||c | c | c ||} 
		\hline
		Description & Variable  & Value \\ [0.5ex] 
		\hline\hline
		Small circle radius & $R_1$ & 47 $m$ \\ 
		Large circle radius & $R_2$ & 95 $m$ \\ 
		Transmission power & $P_{TX}$ & 16 $dBm$ \\
		Noise power & $P_{n}$ & -90 $dBm$ \\
		SNR threshold & $\gamma_{min}$ & 3.9 $dB$ \\
		Bandwidth & $\Delta F$ & 20 $MHz$ \\  
		Payload size & $l$ & 8 $KB$ \\
		Path loss exponent & $\beta$ & 2.6 \\
		SIFS & $SIFS$ & 16 $\mu s$\\
		AIFS & $AIFS$ & 34 $\mu s$\\
		ACK & $T_{ACK}$ & 44 $\mu s$\\
		RTS & $RTS$ & 160 $bit$ \\
		CTS & $CTS$ & 112 $bit$ \\
		Idle slot duration & $\sigma$ & 9 $\mu s$\\
		Minimum contention window & $CW_{min}$ & 16 \\
		Maximum contention window & $CW_{max}$ & 1024 \\ 
		Number of experiments & $n_s$ & 100 \\
		\hline
	\end{tabular}
	\label{tab:parameters}
\end{table}

\section{Numerical Results} 
\label{sec:results}
\begin{figure}[h!]
	\center{\includegraphics[width = 0.97\linewidth]{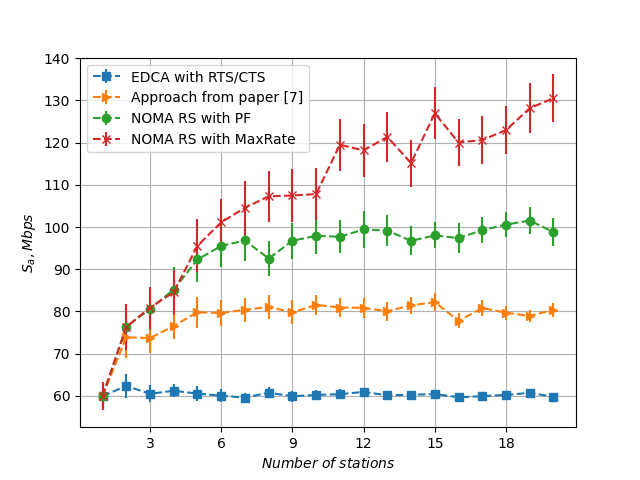}\\ (a)}
	\center{\includegraphics[width = 0.97\linewidth]{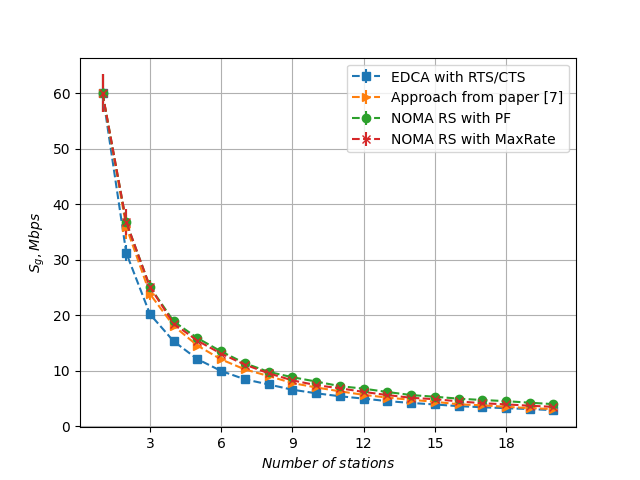} \\ (b)}
	\caption{Aggregated throughput (a) and geometric mean throughput (b) with radius $R_1$.}
	\label{figure:throughput_geomean_47m}
\end{figure}

\begin{figure}[h!]
	\center{\includegraphics[width = 0.97\linewidth]{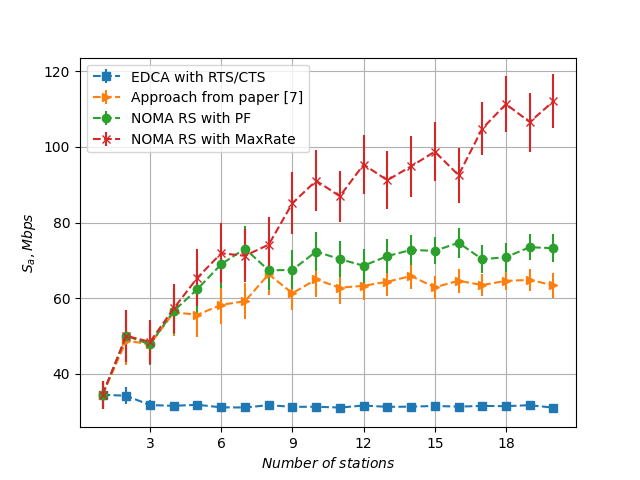}\\ (a)}
	\center{\includegraphics[width = 0.97\linewidth]{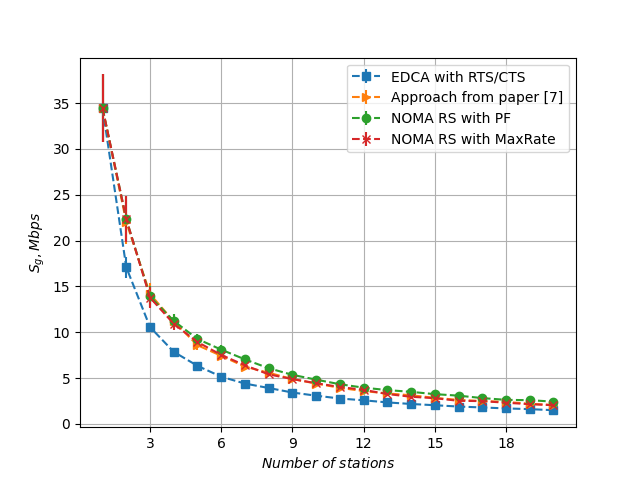} \\ (b)}
	\caption{Aggregated throughput (a) and geometric mean throughput (b) with radius $R_2$.}
	\label{figure:throughput_geomean_95m}
\end{figure}

Let us evaluate the performance of the proposed NOMA RS mechanism in the scenarios described in Section \ref{sec:scenario} with parameters listed in Table \ref{tab:parameters}. For each number of STAs, $n_s$ experiments were carried out with different positions of STAs to average the results.

Figures \ref{figure:throughput_geomean_47m} and \ref{figure:throughput_geomean_95m} display the performance differences between modeling Wi-Fi systems using two radii. In the case of the large radius $R_2$, the overall system performance degrades compared with the results of the system with the small radius $R_1$. It is explained by the appearance of STAs with MCS of low orders and, consequently, transmitting data at the low bitrate. 

Particularly, Figs. \ref{figure:throughput_geomean_47m}a and \ref{figure:throughput_geomean_95m}a represent how the total throughput depends on the number of STAs and different resource schedulers. As can be expected, MaxRate provides the highest aggregated system throughput. 
From the presented Figs. \ref{figure:throughput_geomean_47m}a and \ref{figure:throughput_geomean_95m}a, it is clear that the use of UL-NOMA gives an increase in the aggregated throughput up to 100\% compared with conventional EDCA with RTS/CTS.
Also, the UL-NOMA mechanism outperforms the approach from \cite{korolev2020performance} because the Secondary STA transmits data directly without skipping NOMA slots within the primary frame. It means more payload during one NOMA transmission.

\begin{figure}[ht]
	\center{\includegraphics[width = 0.97\linewidth]{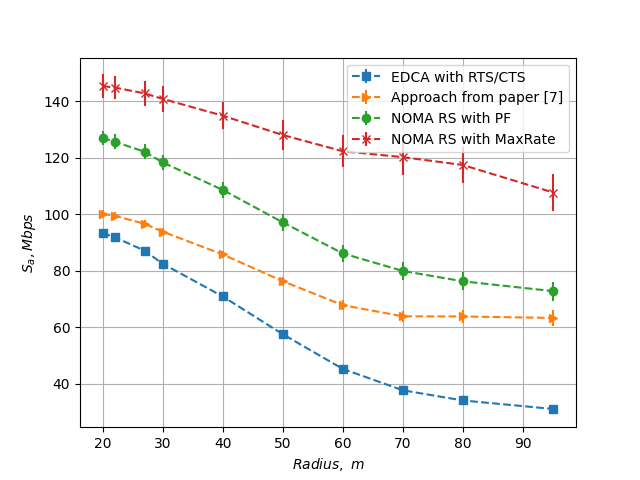}\\ (a)}
	\center{\includegraphics[width = 0.97\linewidth]{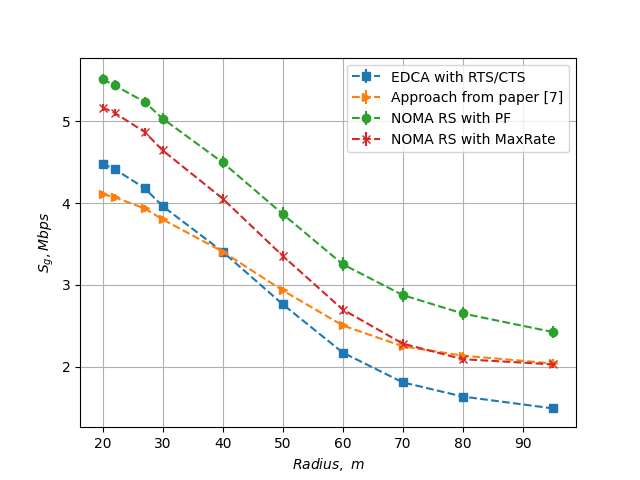} \\ (b)}
	\caption{Aggregated throughput (a) and geometric mean throughput (b) for 20 stations with various radii.}
	\label{figure:throughput_geomean_radius}
\end{figure}

With an increase in the number of STAs in the system, the number of STAs intending to participate in the NOMA transmission also increases. Figs. \ref{figure:throughput_geomean_47m}b and \ref{figure:throughput_geomean_95m}b show that the geometric mean throughput of STAs using EDCA approaches zero faster than with the mechanisms proposed in \cite{korolev2020performance} and described in this paper. A small geometric mean throughput means an unfair distribution of resources within the network. The use of the developed mechanism NOMA RS with PF provides up to 100\% gain in geometric mean throughput due to a proportionally fair resource allocation among Secondary STAs.

Figures \ref{figure:throughput_geomean_radius}a and \ref{figure:throughput_geomean_radius}b show the dependence of the total throughput and the geometric mean on the radius of the system. With an increase in the maximum distance from the farthest STA to the AP, the system performance decreases due to the appearance of STAs with MCS of low orders. Also, in the system under consideration, the problem of hidden stations arises. As a result, a collision probability increases, and overall performance degrades.

\section{Conclusion}
\label{sec:conclusion}
This paper designs the UL-NOMA in Wi-Fi networks and studies its performance. We describe a NOMA RS mechanism that enables synchronous UL-NOMA in Wi-Fi and is compatible with the legacy EDCA mechanism. 

The main idea is that each STA intending to transmit in UL-NOMA sends a unique RS to the AP when this STA detects an RTS frame. If the AP successfully receives this RTS frame and one or several RSs, the AP selects the Secondary STA for UL-NOMA transmission according to some policy. The policy can be the overall rate maximization or proportional fair that maximizes the system's geometric mean throughput. 
After the selection of the Secondary STA, the AP transmits the RS corresponding to the selected Secondary STA along with the CTS frame for the Primary STA. Thus, collisions at the level of UL-NOMA transmissions are completely eliminated.

The results show that the use of UL-NOMA upon the legacy EDCA mechanism gives an increase in the total throughput and geometric mean throughput up to approximately 100\%. This work shows that the use of non-orthogonal multiple access in uplink is promising for further development of Wi-Fi standards.

\bibliographystyle{IEEEtran}
\bibliography{biblio}

\end{document}